\documentclass{webofc}
\usepackage[varg]{txfonts}   % Web of Conferences font
\usepackage{multirow}
\usepackage{lineno}

\begin{document}
\title{\raggedright Physics perspectives with AFTER@LHC \linebreak (A Fixed Target ExpeRiment at LHC)}

\author{\firstname{L.} \lastname{Massacrier}\inst{1}\fnsep\thanks{\email{massacrier@ipno.in2p3.fr}} \and
        \firstname{M.} \lastname{Anselmino}\inst{2} \and
        \firstname{R.} \lastname{Arnaldi}\inst{2} \and
        \firstname{S.J.} \lastname{Brodsky}\inst{3} \and
        \firstname{V.} \lastname{Chambert}\inst{1} \and
        \firstname{C.} \lastname{Da Silva}\inst{4} \and  
        \firstname{J.P.} \lastname{Didelez}\inst{1} \and  
        \firstname{M.G.} \lastname{Echevarria}\inst{5} \and  
        \firstname{E.G.} \lastname{Ferreiro}\inst{6} \and  
        \firstname{F.} \lastname{Fleuret}\inst{7} \and  
        \firstname{Y.} \lastname{Gao}\inst{8} \and  
        \firstname{B.} \lastname{Genolini}\inst{1} \and  
        \firstname{C.} \lastname{Hadjidakis}\inst{1} \and  
        \firstname{I.} \lastname{H\v{r}ivn\'{a}\v{c}ov\'{a}}\inst{1} \and  
        \firstname{D.} \lastname{Kikola}\inst{9} \and 
        \firstname{A.} \lastname{Klein}\inst{4} \and 
        \firstname{A.} \lastname{Kurepin}\inst{10} \and 
        \firstname{A.} \lastname{Kusina}\inst{11} \and 
        \firstname{J.P.} \lastname{Lansberg}\inst{1} \and 
        \firstname{C.} \lastname{Lorc\'e}\inst{12} \and 
        \firstname{F.} \lastname{Lyonnet}\inst{13} \and
        \firstname{G.} \lastname{Martinez}\inst{14} \and
        \firstname{A.} \lastname{Nass}\inst{15} \and
        \firstname{C.} \lastname{Pisano}\inst{16} \and
        \firstname{P.} \lastname{Robbe}\inst{17} \and
        \firstname{I.} \lastname{Schienbein}\inst{18} \and
        \firstname{M.} \lastname{Schlegel}\inst{19} \and
        \firstname{E.} \lastname{Scomparin}\inst{2} \and
        \firstname{J.} \lastname{Seixas}\inst{20} \and
        \firstname{H.S.} \lastname{Shao}\inst{21} \and
        \firstname{A.} \lastname{Signori}\inst{22} \and
        \firstname{E.} \lastname{Steffens}\inst{23} \and
        \firstname{L.} \lastname{Szymanowski}\inst{24} \and
        \firstname{N.} \lastname{Topilskaya}\inst{10} \and
        \firstname{B.} \lastname{Trzeciak}\inst{25} \and
        \firstname{U.I.} \lastname{Uggerh\o j}\inst{26} \and
        \firstname{A.} \lastname{Uras}\inst{27} \and
        \firstname{R.} \lastname{Ulrich}\inst{28} \and
        \firstname{J.} \lastname{Wagner}\inst{24} \and
        \firstname{N.} \lastname{Yamanaka}\inst{1} \and
        \firstname{Z.} \lastname{Yang}\inst{8}       
        \normalsize
        % etc.
}

\institute{ \tiny IPNO, Univ. Paris-Sud, CNRS/IN2P3, Universit\'e Paris-Saclay,  Orsay, France
\and
          Dip. di Fisica and INFN Sez. Torino, Via P. Giuria 1, Torino, Italy
\and
          SLAC National\,Accelerator\,Laboratory, Stanford University, Menlo Park, USA
\and
          LANL, P-25, Los Alamos National Laboratory, Los Alamos, NM 87545, USA
\and
		 INFN Sez. Pavia, Via Bassi 6, 27100 Pavia, Italy 
\and 
		 Dept. de F{\'\i}sica de Part{\'\i}culas, USC, Santiago de Compostella, Spain   
\and 		  
          LLR, \'Ecole Polytechnique, CNRS/IN2P3, Palaiseau, France
\and          
          CHEP, Department of Engineering Physics, Tsinghua University, Beijing, China
\and           
          Faculty of Physics, Warsaw University of Technology,  Warsaw, Poland 
\and      
          Institute for Nuclear Research, Russian Academy of Sciences, Moscow, Russia            
\and 
		Institute of Nuclear Physics Polish Academy of Sciences, PL-31342 Krakow, Poland               
\and   
         CPhT, Ecole Polytechnique, CNRS, Universit\'e Paris-Saclay,  Palaiseau, France
\and    
         Southern Methodist University, Dallas, TX 75275, USA  
\and          
        SUBATECH, IMT Atlantique, Universit\'e de Nantes, CNRS-IN2P3, Nantes, France             
\and          
         Institut f\"ur Kernphysik, Forschungszentrum J\"ulich, J\"ulich, Germany 
\and 
		Dipartimento di Fisica, Universita degli Studi di Pavia, Pavia, Italy         
\and 		
         LAL, Univ. Paris-Sud, CNRS/IN2P3, Universit\'e Paris-Saclay,  Orsay, France
\and           
         LPSC, Universit\'e Grenoble-Alpes, CNRS/IN2P3, 38026 Grenoble, France
\and 
		Institute for Theoretical Physics, T\"ubingen U.,  T\"ubingen, Germany        
\and 	
		LIP and IST, Lisbon, Portugal 	 
\and 
 		Sorbonne Universités, UPMC Univ. Paris 06, UMR 7589, LPTHE, F-75005 Paris, France
CNRS, UMR 7589, LPTHE, F-75005 Paris, France	
\and 
      %  Nikhef and Dept. of Physics and Astronomy, VU  Amsterdam,  Amsterdam, The Netherlands 	
      Theory Center, Thomas Jefferson National Accelerator Facility, 12000 Jefferson Avenue, Newport News, VA 23606, USA
\and 
	   Physics Institute, Friedrich-Alexander University Erlangen-N\"urnberg, Erlangen, Germany        	
\and 	   
       National Centre for Nuclear Research (NCBJ), Hoza 69, 00-681, Warsaw, Poland	   
\and 
       Institute for Subatomic Physics, Utrecht University, Utrecht, The Netherlands     
\and 
	  Department of Physics and Astronomy, University of Aarhus, Denmark           
\and 
      IPNL, Universit\'e Claude Bernard Lyon-I, CNRS/IN2P3, Villeurbanne, France	   
\and 
      Institut f\"ur Kernphysik, Karlsruhe Institute of Technology (KIT), Karlsruhe, Germany
          }

%\linenumbers

\abstract{%
AFTER@LHC is an ambitious fixed-target project in order to address open questions in the domain of proton and neutron spins, Quark Gluon Plasma and high-$x$ physics, at the highest energy ever reached in the fixed-target mode. Indeed, thanks to the highly energetic 7~TeV proton and 2.76~\textit{A}.TeV lead LHC beams, center-of-mass energies as large as $\sqrt{s_{NN}}$~=~115~GeV in pp/pA and $\sqrt{s_{NN}}$~=~72~GeV in AA can be reached, corresponding to an uncharted energy domain between SPS and RHIC. We report two main ways of performing fixed-target collisions at the LHC, both allowing for the usage of one of the existing LHC experiments. In these proceedings, after discussing the projected luminosities considered for one year of data taking at the LHC, we will present a selection of projections for light and heavy-flavour production.  
}
\maketitle

\section{Introduction}

The AFTER@LHC project \cite{AFTERweb} is a proposal to conduct a multi-purpose fixed-target experiment at the LHC by using its highly energetic proton and lead beams. The energy domain which can be probed ranges from $\sqrt{s_{NN}}$~=~72~GeV in AA collisions up to $\sqrt{s_{NN}}$~=~115~GeV in pp/pA collisions, i.e. the largest energies achieved in the fixed-target mode. This mode offers unique opportunities to access the high Feynman-$x_{F}$ domain with high luminosities thanks to the large density of the target. The versatility of the target also allows one to perform studies as a function of the nucleus atomic mass. Finally, depending on the chosen technology, it may be possible to polarise the target. The physics programme of AFTER@LHC has extensively been discussed in Refs \cite{Brodsky:2012vg, Massacrier:2015qba, Kikola:2017hnp, Trzeciak:2017csa} and proposes to address open questions in the domain of high-$x$, Spin and Quark Gluon Plasma (QGP) physics. In the high-$x$ programme, the aim is to advance our understanding of the high-$x$ gluon, antiquark and heavy-quark content in the nucleon and in the nucleus: by, for instance, constraining the quark Parton Distribution Functions (PDF) and nuclear PDFs (in the EMC region) with Drell-Yan measurements; by searching for the existence of a possible non-perturbative source of $c$ or $b$ quarks in the proton, which is an important input for high-energy-neutrino and cosmic-ray physics; by looking for $W$ boson production near threshold to constrain the light quark sea PDFs at large-$x$.  \newline %\cite{Brodsky:2012vg, Lansberg:2016urh, Lansberg:2012kf, Lansberg:2012wj, Rakotozafindrabe:2012ei, Lorce:2012rn, Lansberg:2012sq, Rakotozafindrabe:2013au, Lansberg:2013wpx, Rakotozafindrabe:2013cmt, Lansberg:2014myg, Massacrier:2015nsm, Massacrier:2015qba, Lansberg:2016gwm, Kikola:2017hnp, Trzeciak:2017csa} and proposes to address open questions in the domain of high-$x$, Spin and Quark Gluon Plasma (QGP) physics. In the high-$x$ programme, the aim is to advance our understanding of the high-$x$ gluon, antiquark and heavy-quark content in the nucleon and in the nucleus: by, for instance, constraining the quark Parton Distribution Functions (PDF) and nuclear PDFs (in the EMC region) with Drell-Yan measurements; by searching for the existence of a possible non-perturbative source of $c$ or $b$ quarks in the proton, which is an important input for high-energy-neutrino and cosmic-ray physics; by looking for $W$ boson production near threshold to constrain the light quark sea PDFs at large-$x$.  \newline
\indent{In the Spin physics programme (not discussed here),  the goal is to advance our understanding of the dynamics and spin of quarks and gluons inside polarised (and unpolarised) nucleons, in particular the Orbital Angular Momentum of quarks and gluons in the proton.} \newline
\indent{The final part of the AFTER@LHC physics programme concerns the study of heavy-ion collisions toward large rapidities. In Pb-A collisions, at $\sqrt{s_{NN}}$ = 72 GeV, AFTER@LHC is probing the region of high temperature ($\sim$ 1.5 T$_{c}$) and low baryon chemical potential, where QGP formation is expected to occur. At such a temperature, the $\Upsilon(2S,3S)$ excited states are expected to be suppressed in the QGP \cite{Mocsy:2008eg}, thus allowing the calibration of its temperature. Conducting measurements of various quarkonium states (together with open heavy-flavours (HF)) as a function of rapidity and system size would permit to scan the phase-transition region. Moreover, AFTER@LHC can study the transport properties of the QGP accounting for its longitudinal expansion. By measuring particle yields and flow coefficients v$_{N}$ as a function of rapidity, AFTER@LHC can access the temperature dependence of the medium shear viscosity while probing different energy densities. A proper interpretation of the AA data probably requires a complete set of v$_{N}$ coefficient measurements in smaller systems (pp, pA) to study the collectivity with new observables, like heavy-flavour hadrons, which will be abundantly produced at AFTER@LHC. Other items of interest also include the study of the heavy-quark energy-loss mechanisms in the QGP via $D$ meson measurements as a function of the rapidity and p$_{\rm T}$, and the universality of the initial-state effects from pA to AA collisions with the Drell-Yan probe. In these proceedings, we will focus on the performance of AFTER@LHC for light and heavy-flavour production. }

\section{Possible technical implementations at the LHC and luminosities}

In the following, we will discuss the usage of an internal (solid or gaseous) target inside one of the existing LHC experiments (LHCb or ALICE) as a main way to achieve the AFTER@LHC physics goals. Such kinds of solutions enable to conduct a fixed-target programme at a limited cost, with a limited civil engineering, and on a shorter timescale. The LHCb experiment has demonstrated the feasibility of injecting low-density noble gases inside the vacuum chamber of its Vertex Locator detector using the SMOG system \cite{Aaij:2014ida}. This setup has proven that a parasitic operation of a fixed-target programme with a collider programme is possible, without inducing a decrease of the beam lifetime. However, the physics reach with such a system is limited by: the low gas pressure; the choice of gas species; the limited running time due to the absence of dedicated pumping systems close to the interaction point; the absence of target polarisation. Typical luminosities collected in 2015 in pAr collisions, during about 17 hours of data taking amount to few nb$^{-1}$ \cite{LHCb:2017qap}. To further increase the gas pressure, two setups are under study. An internal gas-jet is currently used at the RHIC collider to measure proton beam polarisation \cite{Zelenski:2005mz}, while a storage-cell gas target has been used by the Hermes experiment at the Hera collider \cite{Airapetian:2004yf}. With both systems, target areal density as high as 10$^{15}$ - 10$^{16}$~H$_{2}$.cm$^{-2}$ could be reached with unpolarised H$_{2}$, i.e. an increase of the target areal density by several orders of magnitude with respect to the SMOG system. Thanks to a powerful differential pumping system maintaining a high local gas density in the target region, polarised hydrogen target areal density about two orders of magnitude larger than the internal gas-jet one can be obtained with the storage-cell gas target. Finally, the usage of an internal solid target in combination with a bent crystal deflecting the beam halo is also under study \cite{Scandale:2010zzb}. Beam fluxes on the order of  5 $\times 10^{8}$ p/s and 10$^{5}$ Pb/s can be extracted by means of a bent crystal. An unpolarised solid target, with a length of 5 mm along the beam direction, is currently considered. Such a length should allow an installation inside the LHC beam pipe and limited multiple-scattering inside the target. Table~\ref{tab-1} shows the target areal density, instantaneous and integrated luminosities over a year for the gas-jet target, storage-cell target and a bent crystal coupled to a solid-target. Only a selection of beam and target type combinations relevant for the discussions and results in these proceedings is shown. The largest integrated luminosities per year in pH$^{\uparrow}$ ($\sim$ 10 fb$^{-1}$) and PbXe ($\sim$ 30 nb$^{-1}$) collisions are obtained with the storage-cell. These luminosities should be considered as maximal and can be further decreased because of the detector data acquision rate capabilities (e.g. the ALICE detector case) or a decrease of the beam lifetime (a 15$\%$ beam consumption over a fill has been assumed for PbXe collisions).

%To further increase the gas pressure, two setups are under study. An internal gas-jet is currently used at the RHIC collider to measure proton beam polarisation \cite{Zelenski:2005mz}. It uses a polarised free-atomic-beam source (ABS) made of nine vacuum chambers providing a system with nine stages of differential pumping. A hydrogen flux as high as 10$^{15}$ - 10$^{16}$ H$_{2}$/s (first estimates) could be reached with unpolarised H$_{2}$, leading to an increase of the target areal density by several orders of magnitude with respect to the SMOG system, in a much smaller location. The storage-cell gas target has been used by the Hermes experiment at the Hera collider \cite{Airapetian:2004yf}. It is constituted by a long straight openable tube which would be placed inside the vacuum of the LHC beam pipe \cite{Steffens:2015kvp}. A feed tube, coupled to an ABS system permits the injection of polarised gases. A simple capillary can be used as well to inject unpolarised gases. A powerful differential pumping system is required in order to maintain a high local gas density in the target region (about two orders of magnitude larger than the one obtained with an internal polarised hydrogen gas jet) without affecting the global LHC vacuum. 

\begin{table}[!htpb]
\centering
\caption{\footnotesize Summary table of the target areal density, instantaneous and integrated luminosities over a year, for the various technical solutions described in the text, and a selection of beam and target types. The solid target is considered to be 5mm-thick along the beam direction. The proton (lead) LHC year is assumed to last 10$^{7}$ s (10$^{6}$ s) respectively. The proton (lead) beam flux is considered to be 3.63 $\times$ 10$^{18}$ s$^{-1}$ (4.66 $\times$ 10$^{14}$ s$^{-1}$)  for the gas-jet and storage-cell solutions. The symbol $\uparrow$ indicates that the target is polarised.}
\label{tab-1}       % Give a unique label
%% For LaTeX tables you can use
\resizebox{\textwidth}{!}{%
%\tiny
\begin{tabular}{|c|c|c|c|c|c|}
\hline
Technical Solution    										& Beam type & Target type 					& $\theta_{\rm target}$						&	 $\cal{L}$ 			& $\cal{L}_{\rm int}$   \\ 
							 										& 				    & 				   					 &      (cm$^{-2}$)				 					& (cm$^{-2}$.s$^{-1}$) & (pb$^{-1}$/year)   \\\hline
\multirow{4}{*}{Gas-Jet Target}							& p 				&   H$\uparrow$	 			&		   1.2 $\times$ 10$^{12}$			&		4.3 $\times$ 10$^{30}$	&	 43					 \\\cline{2-6}
																	& p 				&   H$_{2}$					&		   10$^{15}$  -  10$^{16}$		&		3.6 $\times$ 10$^{33}$ - 3.6 $\times$  10$^{34}$	&	 36 $\times$ 10$^{3}$ -  36 $\times$ 10$^{4}$	\\\cline{2-6}
																	& Pb				&	 H$\uparrow$				&		   1.2 $\times$ 10$^{12}$			&		5.6 $\times$ 10$^{26}$	&	 0.56 $\times$ 10$^{-3}$		 \\\cline{2-6}
																	& Pb 				&   H$_{2}$	 			     &		   10$^{15}$  -  10$^{16}$		&		4.7 $\times$ 10$^{29}$ - 4.7 $\times$ 10$^{30}$	&	 0.47 - 4.7 						
 \\\hline
\multirow{4}{*}{Storage-Cell Target}    				&  p				&	 H$\uparrow$				&			2.5 $\times$ 10$^{14}$		    &		9.2 $\times$ 10$^{32}$	&	 9200				 \\\cline{2-6}	
																	&  p				&	 Xe							&			6.4 $\times$ 10$^{13}$			&		2.3 $\times$ 10$^{32}$	&	 2300				\\\cline{2-6}		
																	&  Pb				&	 H$\uparrow$				&			2.5 $\times$ 10$^{14}$		    &		1.2 $\times$ 10$^{29}$	&	0.120				\\\cline{2-6}
																	&  Pb				&	 Xe							&			6.4 $\times$ 10$^{13}$			&		3.0 $\times$ 10$^{28}$	&	0.030				\\\hline	
\multirow{2}{*}{Bent Crystal + Solid Target}		&	p				&	Pb							&			1.6 $\times$ 10$^{22}$			&		8.2 $\times$	10$^{30}$ &	    82					  \\\cline{2-6}
																	&  Pb				&	Pb							&			1.6 $\times$ 10$^{22}$			&		1.6 $\times$ 10$^{27}$	&  1.6 $\times$ 10$^{-3}$
 \\\hline
\end{tabular}}
%% Or use
%\vspace*{5cm}  % with the correct table height
\end{table}

The ALICE and LHCb detectors are both well suited to conduct a fixed-target programme of AFTER@LHC. The LHCb detector is fully instrumented in the forward region, with excellent particle identification (PID) performance. It has a mid- to backward coverage in the center-of-mass frame in the fixed-target mode, which reaches large negative Feynman x$_{F}$. The ALICE detector also provides a  similar coverage in the mid- to backward region thanks to its current muon arm and future Muon Forward Tracker. The long absorber upstream the muon tracking stations is a key feature to reduce the combinatorial background for Drell-Yan studies in pA/AA. In addition, the central barrel of ALICE provides a very backward coverage in the fixed-target mode with good PID capabilities, allowing one to reach the end of the phase space for several identified soft probes. 

\section{Projected performance for light and heavy-flavours}

\begin{figure*}[!htpb]
\centering
\includegraphics[width=11cm,clip]{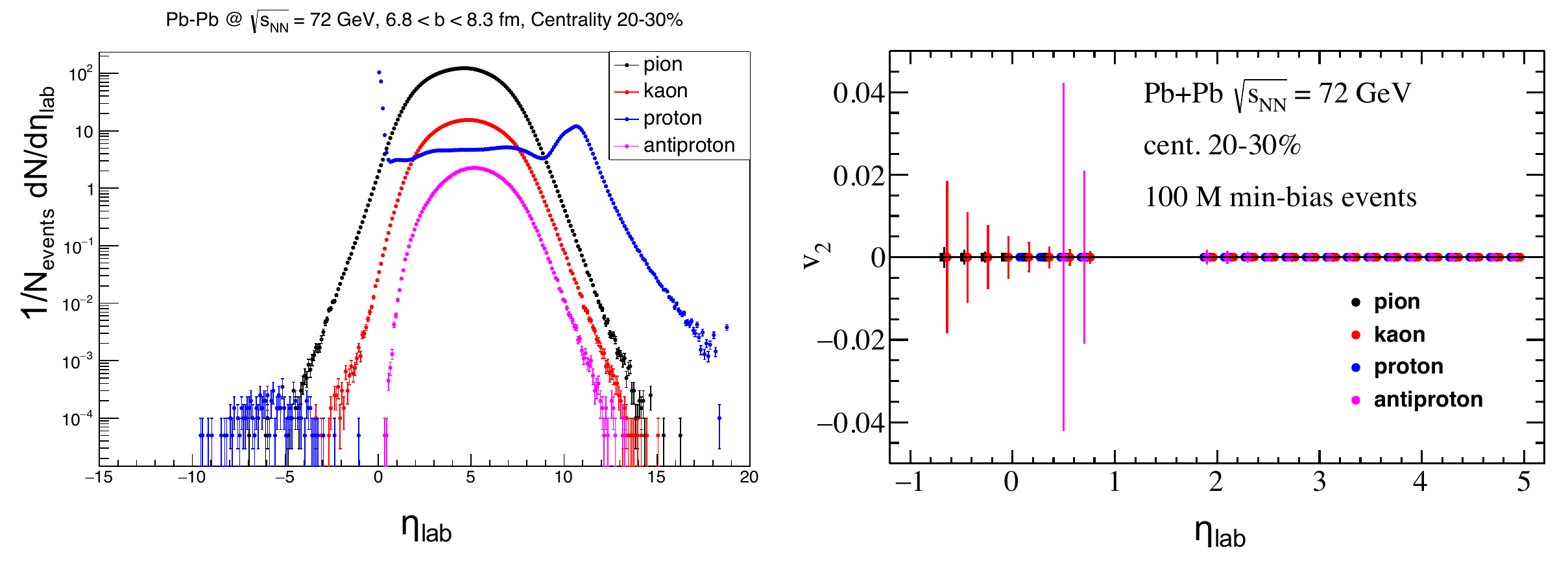}
 %\includegraphics[width=7cm,clip]{IdPart_PbPb}
 %\includegraphics[width=7cm,clip]{v2_IdPart_PbPb}
% to insert the figure file. (like the \includegraphics command above) 
% and remove the following \vspace line:
%\vspace*{5cm}       % Give the correct figure height in cm

\caption{\footnotesize Light-particle multiplicity distributions (left) and elliptic flow $v_{2}$ (right) as a function of $\eta_{\rm lab}$ for fixed-target Pb-Pb collisions, in the centrality range 20-30\%. Pions (kaons,  protons and antiprotons) are represented as black (red, blue, pink) dots, respectively. In the right panel, the statistical uncertainties on the $v_{2}$ coefficients (vertical uncertainty bars) have been computed in the ALICE central-barrel and LHCb detector acceptances, assuming the collection of 10$^{8}$ minimum bias events.}

\label{fig-2}       % Give a unique label
\end{figure*}

The measurement of the flow coefficients at mid- and backward rapidity in the c.m.s frame is interesting to test hydrodynamic calculations accounting for the longitudinal expansion of the formed medium. Indeed, AFTER@LHC would give us insight on the QGP behaviour at low energy, in a rapidity region seldom probed. Figure \ref{fig-2} (left) shows the multiplicity distribution of identified particles (pion, kaon, proton, antiproton) as a function of pseudo-rapidity in the lab ($\eta_{\rm lab}$), in Pb-Pb collisions at $\sqrt{s_{NN}}$~=~72~GeV, in the centrality range 20-30$\%$, generated with the EPOS@LHC Monte Carlo (MC) generator \cite{Pierog:2013ria,Werner:2005jf}. Identified light particles can be studied in a wide rapidity range with the LHCb and ALICE detectors, and down to their phase space limit with the ALICE central barrel. Based on the identified-particle yields from the figure \ref{fig-2} (left), the projections of the statistical uncertainties on the elliptic flow measurement $v_{2}$, as a function of $\eta_{\rm lab}$ for ALICE and LHCb detector acceptances, are shown in figure \ref{fig-2} (right). A minimum particle p$_{\rm T}$ of 0.2 GeV/c (ALICE) and 0.5 GeV/c (LHCb) is required for the PID. The elliptic-flow of identified-particles can be accurately measured in LHCb, after only few hours of Pb-Pb data taking with a solid target (10$^{8}$ minimum bias events). In the ALICE central-barrel acceptance, despite the lower yields, an absolute statistical uncertainty on $v_{2}$ of 4$\%$ for antiprotons, $2\%$ for kaons and better than 1$\%$ for pions and protons can already be reached. 

%
%%\begin{figure*}[!htpb]
%\centering
%\includegraphics[width=12.5cm,clip]{Fig1}
% %\includegraphics[width=7cm,clip]{ppmult}
% %\includegraphics[width=7cm,clip]{AAmult}
%% Use the relevant command for your figure-insertion program
%% to insert the figure file. (like the \includegraphics command above) 
%% and remove the following \vspace line:
%%\vspace*{5cm}       % Give the correct figure height in cm
%
%\caption{Charged particles multiplicity distributions as a function of pseudo-rapidity in the lab for pp collisions generated with Pythia8 MC (left) and in AA collisions generated with EPOS MC (right). On the left panel, fixed target pp collisions at $\sqrt{s}$ = 72 GeV (blue) and $\sqrt{s}$ = 115 GeV (red) are shown. On the right panel, fixed target Pb-Ar (black dashed line), Pb-Xe (green dashed line) and Pb-Pb (red dashed line) collisions in the centrality range 0-10$\%$ at $\sqrt{s_{NN}}$ = 72 GeV are shown together with Pb-Pb collisions in collider mode at $\sqrt{s_{NN}}$ = 5.5 TeV, in the centrality range 0-10$\%$ (blue line) and 40-50$\%$ (pink line).}
%
%
%\label{fig-1}       % Give a unique label
%\end{figure*}

\begin{figure*}[!htpb]
\centering
 \includegraphics[width=11cm,clip]{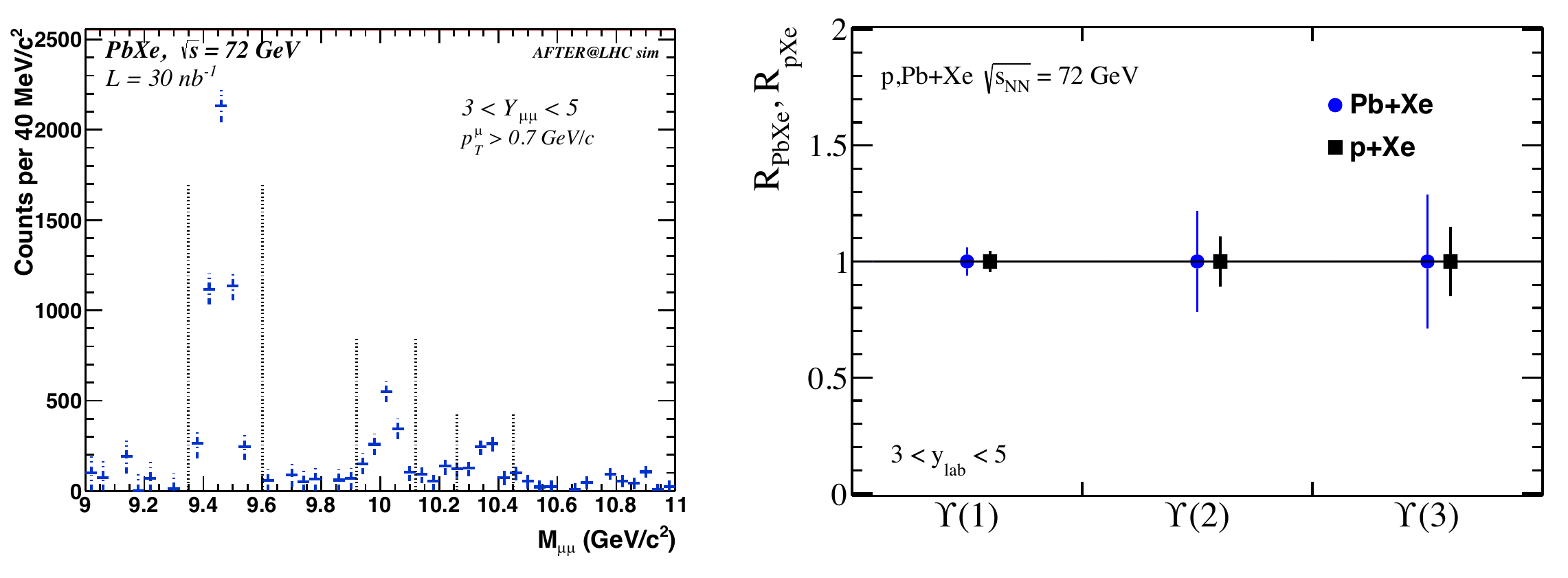}
% to insert the figure file. (like the \includegraphics command above) 
% and remove the following \vspace line:
%\vspace*{5cm}       % Give the correct figure height in cm

\caption{\footnotesize Left panel: Dimuon invariant-mass distribution in the $\Upsilon(nS)$ region after combinatorial background subtraction, in Pb-Xe collisions at $\sqrt{s_{NN}}$ = 72 GeV ($\cal{L}_{\rm int}$~=~30 nb$^{-1}$). Single muons are required to have $p_{T} >$~0.7~GeV/c and dimuons to be within 3 $< y_{\rm lab} <$ 5. Right panel: Projection of the statistical uncertainties (vertical bars) on the nuclear modification factor in pXe collisions (black square), Pb-Xe collisions (blue dot) for $\Upsilon(1S)$, $\Upsilon(2S)$ and $\Upsilon(3S)$, in the rapidity range 3 $< y_{\rm lab} <$ 5.}

\label{fig-3}       % Give a unique label
\end{figure*}

AFTER@LHC will also be well suited to measure several quarkonium states in order to assess the thermodynamical properties of the QGP in AA collisions. In particular, precise measurements of bottomonium states in pp, pA and AA collisions as a function p$_{\rm T}$, rapidity and system size will provide a range of densities and temperatures to identify the conditions for deconfinement and the quantification of the Cold Nuclear Matter (CNM) effects. %the search of the deconfinement phase transition, the calibration of the QGP temperature (the recombination of $b\bar{b}$ pairs is negligible) and the quantification of the Cold Nuclear Matter (CNM) effects. 
Figure \ref{fig-3} (left) shows the dimuon invariant mass distribution for $\Upsilon$(nS) states after combinatorial background subtraction (using dimuon-like sign pairs), in Pb-Xe collisions at $\sqrt{s_{NN}}$~=~72~GeV, generated with Pythia8~\cite{Sjostrand:2007gs} and HELAC-Onia~\cite{Shao:2012iz,Shao:2015vga} MC generators\footnote{pp simulations were scaled to PbXe collisions assuming no nuclear effects (N$_{\rm coll}$ scaling for the hard probes).}. The performance of an LHCb-like detector with a storage-cell target have been assumed ($\cal{L}_{\rm int}$~=~30 nb$^{-1}$). Figure \ref{fig-3} (left) is an example of the $\Upsilon$(nS) yields which can be collected in the rapidity range 3~$< y_{\rm lab} <$~5. The three upsilon states are clearly visible (about 450 $\Upsilon$(3S) are expected per year if no nuclear effects are considered). Figure \ref{fig-3} (right) represents the corresponding projection of the statistical precision on the nuclear modification factor in pXe collisions (R$_{\rm pXe}$, black squares) and Pb-Xe collisions (R$_{\rm PbXe}$, blue dots) for the three upsilon states, in the rapidity range 3~$< y_{\rm lab} <$5. A total sampled luminosity of $\cal{L}_{\rm int}$~=~250 pb$^{-1}$ in pXe, and $\cal{L}_{\rm int}$~=~2~pb$^{-1}$ in pp collisions have been assumed. A 7$\%$ (30$\%$) statistical uncertainty on the R$_{\rm PbXe}$ measurement for $\Upsilon$(1S) ($\Upsilon$(3S)) respectively is within reach, allowing for the study of $\Upsilon$ excited state suppression in the QGP in a completely new energy domain. The statistical uncertainty of 5$\%$ (15$\%$) on R$_{\rm pXe}$ for $\Upsilon$(1S) ($\Upsilon$(3S)) respectively will further constrain CNM effects and help to interpret the AA data. 

\begin{figure*}[!htpb]
\centering
 \includegraphics[width=10cm,clip]{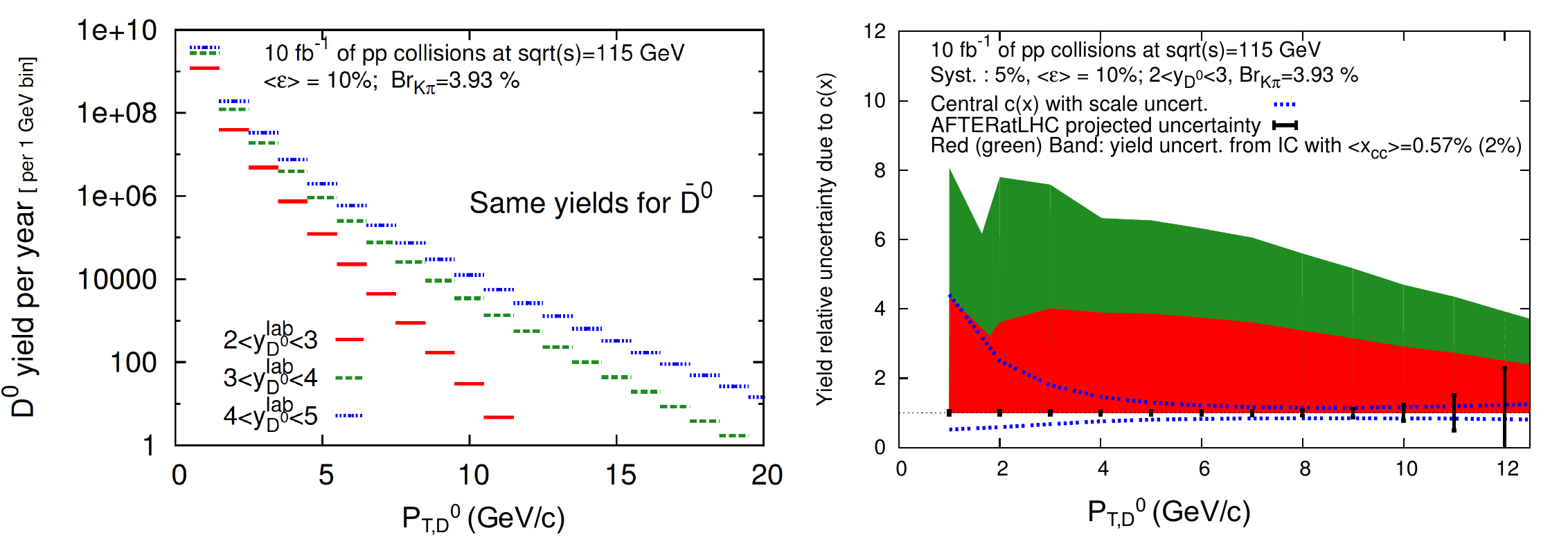}
% to insert the figure file. (like the \includegraphics command above) 
% and remove the following \vspace line:
%\vspace*{5cm}       % Give the correct figure height in cm

\caption{\footnotesize Left panel: D$^{0}$ meson yield per year as a function of $p_{\rm T}$ in fixed-target pp collisions at $\sqrt{s}$~=~115~GeV, for three D$^{0}$ rapidity ranges: 2 $< y_{lab} <$ 3 (red line), 3 $< y_{lab} <$ 4 (green dashed line), 4 $< y_{lab} <$ 5 (blue dotted line). Right panel: Relative uncertainty on the D$^{0}$ meson yield due to the IC content of the proton as a function of $p_{\rm T}$. The vertical black lines correspond to the statistical projections of the AFTER@LHC yearly D$^{0}$ meson yield in pp collisions at $\sqrt{s}$~=~115 GeV, in 2 $< y_{lab} <$ 3.} %add description for the coulours?

\label{fig-4}       % Give a unique label
\end{figure*}

%AFTER@LHC is also expected to perform a complete set of open heavy-flavour studies.
In order to better understand the yield of neutrinos originating from charmed hadrons produced during the collision of ultra-high energy cosmic rays with the earth atmosphere, the charm-hadroproduction cross section needs to be accurately determined. In particular, the presence of a non-perturbative intrinsic charm (IC) component in the proton can alter the charm hadron yields.  %The contribution from IC in the proton could therefore have a significant impact on the flux of neutrinos from charm at high energy.
 AFTER@LHC, covering the large negative Feynman-$x_{F}$ region down to very low p$_{\rm T}$ for charmed hadrons is ideal to put constraints on the IC component of the proton and therefore on the flux of neutrinos from charm at high energy. Figure \ref{fig-4} (left) represents the D$^{0}$ meson yield as a function of p$_{\rm T}$, in pp collisions at $\sqrt{s}$ = 115 GeV, for one year of data taking assuming a LHCb-like detector with a storage-cell target. Large yields up to about 10 GeV/c are expected in the 3 rapidity ranges considered. In figure \ref{fig-4} right, the impact of the IC on the relative D$^{0}$ yield uncertainty is shown as a function of p$_{\rm T}$ in the rapidity range 2 $< y_{lab} <$ 3 and compared with the projected uncertainties on the D$^{0}$ yields predicted for AFTER@LHC (black lines). The red (green) band assumes a fraction of 0.57$\%$ (2$\%$) of IC and are derived from theoretical cross sections from Ref \cite{Brodsky:2015fna}. Even considering a 5$\%$ systematical uncertainty on the D$^{0}$ yield measurement at AFTER@LHC, the precision reached for p$_{\rm T}$ <12 GeV/c will permit to set strong constraints on IC models.

%neutrino from charm flux
%word on pA and AA measurements also at reach. 

\section{Conclusion}
The AFTER@LHC study group is currently investigating the possibilities offered  by a future multi-purpose fixed-target experiment at the LHC, in the sector of high-$x$, spin and heavy-ion physics. Several possible implementations of a fixed-target setup inside an existing LHC experiment (ALICE or LHCb) are currently explored and the expected integrated yearly luminosities have been derived for each setup. In these proceedings, we have presented a selection of projected performance for light and heavy-flavour production at AFTER@LHC. It has been shown that the elliptic flow of identified particles can be measured accurately with an ALICE- or LHCb-like detector, after only few hours of  Pb-Pb data taking at $\sqrt{s_{NN}}$~=~72 GeV, opening the path for a study of the temperature dependence of the medium shear viscosity. Both in pXe, and PbXe collisions, the $\Upsilon(nS)$ yields can already be measured after one year of data taking. It will allow to calibrate the QGP thermometer in an energy domain between the ones probed at SPS and RHIC using  $\Upsilon$ excited states suppression in the QGP. Accurate measurements in pXe collisions will complement the PbXe studies, for a proper interpretation of the CNM effects. Finally, the large D$^{0}$ meson yields expected to be recorded in pp collisions at $\sqrt{s}$~=~115~GeV at large-$x$, will be decisive to constrain the proton IC content and to provide valuable inputs for cosmic-ray physics. The study of open heavy-flavour hadrons in pA/AA will permit to study collectivity in small systems with new probes and to study heavy-quark energy-loss mechanisms in a dense medium.

\footnotesize

\section{Acknowledgement}

This research was supported by the French P2IO Excellence Laboratory, the French CNRS via the grants FCPPL-Quarkonium4AFTER $\&$ D\'efi Inphyniti–Th\'eorie LHC France, the P2I department of the Paris-Saclay University and by the COPIN-IN2P3 Agreement. AS acknowledges support from U.S. Department of Energy contract DE-AC05-06OR23177, under which Jefferson Science Associates, LLC, manages and operates Jefferson Lab.

%
%\section{Introduction}
%\label{intro}
%Your text comes here. Separate text sections with
%\section{Section title}
%\label{sec-1}
%For bibliography use \cite{RefJ}
%\subsection{Subsection title}
%\label{sec-2}
%Don't forget to give each section, subsection, subsubsection, and
%paragraph a unique label (see Sect.~\ref{sec-1}).

%For one-column wide figures use syntax of figure~\ref{fig-1}
%\begin{figure}[h]
%\centering
%\includegraphics[width=1cm,clip]{tiger}
%\label{fig-1}       % Give a unique label
%\end{figure}

%For two-column wide figures use syntax of figure~\ref{fig-2}
%\begin{figure*}
\centering
%\vspace*{5cm}       % Give the correct figure height in cm
%\label{fig-2}       % Give a unique label
%\end{figure*}

%For figure with sidecaption legend use syntax of figure
%\begin{figure}
%\centering
%\sidecaption
%\includegraphics[width=5cm,clip]{tiger}
%\label{fig-3}       % Give a unique label
%\end{figure}

%For tables use syntax in table~\ref{tab-1}.
%\begin{table}
%\centering
%\label{tab-1}       % Give a unique label
%% For LaTeX tables you can use
%\begin{tabular}{lll}
%\hline
%first & second & third  \\\hline
%number & number & number \\\hline
%\end{tabular}
%% Or use
%\vspace*{5cm}  % with the correct table height
%\end{table}
%
% BibTeX or Biber users please use (the style is already called in the class, ensure that the "woc.bst" style is in your local directory)
% \bibliography{name or your bibliography database}
%\begin{thebibliography}{}
%Journal Author, Journal \textbf{Volume}, page numbers (year)
%Book Author, \textit{Book title} (Publisher, place, year) page numbers
% etc
%\end{thebibliography}
%\bibliographystyle{utphys_cp}

\bibliography{template}

\end{document}